\newcommand{\ignore}[1]{}

\documentstyle[mprocl]{article}

\input{psfig}
\def\be{\begin{equation}}
\def\ee{\end{equation}}
\def\bea{\begin{eqnarray}}
\def\eea{\end{eqnarray}}

\def\bmat{           \left |  \begin{array}{cc} }
\def\emat{ \end{array} \right |   }    
\newcommand\half{{\textstyle{1\over2}}}

\newcommand{\smfrac}[2]{{\textstyle{{#1}\over{#2}}}}

\newcommand{\E}[1]{e^{\textstyle {#1}}}

\newcommand{\da}{^\dagger}

\begin{document}

\title{THE QCD  ABACUS:\\ A New Formulation for Lattice Gauge  Theories~\footnote{Lecture  at "APCTP-ICTP Joint International Conference '97  on Recent Developments in Non-perturbative Method" May, 1997, Seoul,  
Korea. MIT Preprint CTP 2693. }}

\author{ RICHARD  C. BROWER }

\address{Department of Physics, Boston University, \\ 590 Commonwealth Avenue,
Boston, MA 02215,  USA  \\E-mail: brower@ctp.mit.edu}


\maketitle\abstracts{ A quantum Hamiltonian is constructed for SU(3)
lattice QCD entirely from color triplet Fermions --- the standard
quarks and a new Fermionic ``constituent'' of the gluon we call
``rishons''. The quarks are represented by Dirac spinors on each site
and the gauge fields by rishon-antirishon bilinears on each link
which together with the local gauge transforms are the generators of
an SU(6) algebra. The effective Lagrangian for the path integral lives
in $R^4 \times S^1$ Euclidean space with a compact ``fifth time'' of
circumference ($\beta$) and non-Abelian charge ($e^2$) both of which
carry dimensions of length. For large $\beta$, it is conjectured that
continuum QCD is reached and that the dimensionless ratio $g^2 =
e^2/\beta$ becomes the QCD gauge coupling.  The quarks are introduced
as Kaplan chiral Fermions at either end of the finite slab in fifth
time. This talk will emphasize the gauge and algebraic structure of
the rishon or link Fermions and the special properties that may lead
to fast discrete dynamics for numerical simulations and new
theoretical insight. }

\section{Introduction}

A quarter of a century after its discovery, solving QCD remains one of
the major challenges for theoretical particle physics. While Quantum
Chromodynamics is generally acknowledged to be a complete theory of all
hadronic or nuclear interactions, only special perturbative
consequences are well understood.  In 1974 a non-perturbative
formulation of QCD was given by Wilson~\cite{Wil74} using a lattice
regulator which maps the quantum field theory onto a classical
statistical mechanics problem in a 4-d Euclidean space.  Although a
variety of techniques have been developed to solve lattice field
theories, at present the most powerful tool is the Monte Carlo
sampling of the partition function of the corresponding classical
statistical mechanics system. Moreover, the most efficient
numerical algorithms for QCD suffer from critical slowing down when
the continuum limit is approached and thus overwhelm even the fastest 
supercomputers. Consequently it is reasonable to look for new
formulation of the QCD problem. This talk will present a radically
new approach based on recent work by Brower, Chandrasekharan and
Wiese~\cite{BroQCD}. Additional background material for this approach
has already been given by Uwe-Jens Wiese in his talk~\cite{WieseTalk} 
just preceding this one.

\begin{figure}
\psfig{figure=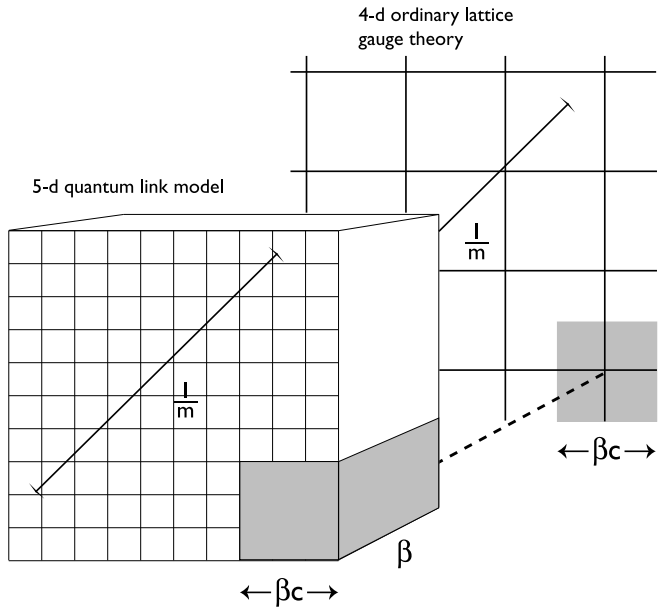,height=80.0mm}
\caption{Dimensional reduction in slab geometry: An effective
4-d lattice theory results with lattice spacing $\beta c$ and a
gluonic correlation length $1/m$ which is exponential in $\beta$. 
\label{fig:dimred}}
\end{figure}

Here we present our alternative non-perturbative approach to QCD in
the framework of quantum link models, which leads to a new
computational framework we call the QCD Abacus~\footnote{Abacus: an
instrument for performing calculations by sliding counters along
rods. From Greek {\em abax} ($\alpha \beta \alpha \xi$), literally
slab.}.  Such gauge models were first discussed by Horn ~\cite{Hor81},
and studied in more detail by Orland and Rohrlich ~\cite{Orl90} and by
Chandrasekharan and Wiese~\cite{Cha96}.  In these models the 4-d
classical statistical mechanics problem of standard lattice gauge
theory is replaced by a problem of 4-d quantum statistical
mechanics. In particular, the classical Euclidean action is replaced
by a Hamilton operator. As a consequence, the ordinary c-numbers in
the standard formulation of lattice gauge theory for the gauge
matrices and their local gauge transformations are replaced by
non-commuting operators acting in a Hilbert space.  Thus one may say
that the geometry of the gauge manifold is replaced by a
non-commutative geometry. Nonetheless as discussed in detail in
Wiese's talk, both 4-d non-Abelian quantum link gauge theories and 2-d
quantum sigma models are expected to ``dimensionally reduce'' to an
effective continuum field theory exhibiting asymptotically free
scaling as the extra compact dimension becomes large (see
Fig.\ref{fig:dimred}).
These crucial physical motivations will for the most part not be
repeated here so the reader is referred to the
literature~\cite{Cha96,BroQCD,WieseTalk} for them.

In spite of these interesting developments, until the recent paper by
Brower, Chandrasekharan and Wiese~\cite{BroQCD}, no candidate quantum
link theory for QCD itself had been proposed due to the difficulty in
finding a mechanism to break the U(3) gauge group down to SU(3).  All
earlier constructions suffered from an extra U(1) gauge symmetry with
a strongly coupled ``photon'' mode which made any serious application
to QCD unlikely.  Here we describe the algebraic insight that allowed
us to construct a candidate quantum link Hamiltonian for QCD.  The
structure of our proposed QCD quantum link Hamiltonian is elegantly
expressed in terms of new Fermionic triplet/anti-triplet fields ($3,
\bar 3$) which in a tight binding mode form ``composite'' gluonic
degrees of freedom.  We call these Fermionic constituents of the
gauge links, ``rishons'' or link Fermions~\cite{BroLinkFermion}.  We
believe this quantum link formulation of non-perturbative QCD is both correct and
compelling for its simplicity. Perhaps it will lead to more powerful
methods for attacking long-standing problems in lattice gauge theories
by analytical and/or numerical methods.

This talk is organized as follows. In section 2 quantum link models
are formulated in terms of rishons --- the Fermionic constituents of
the gluons. The inclusion of quarks is discussed in section 3 within
the framework of Shamir's variant of Kaplan's Fermion method. Section
4 contains a discussion of the prospects for deeper theoretical
understanding, an analytical approach to the large N limit, for
example, and cluster methods for Monte Carlo simulations of quantum
link gauge theories.

\section{The QCD Abacus}

\begin{figure}
\psfig{figure=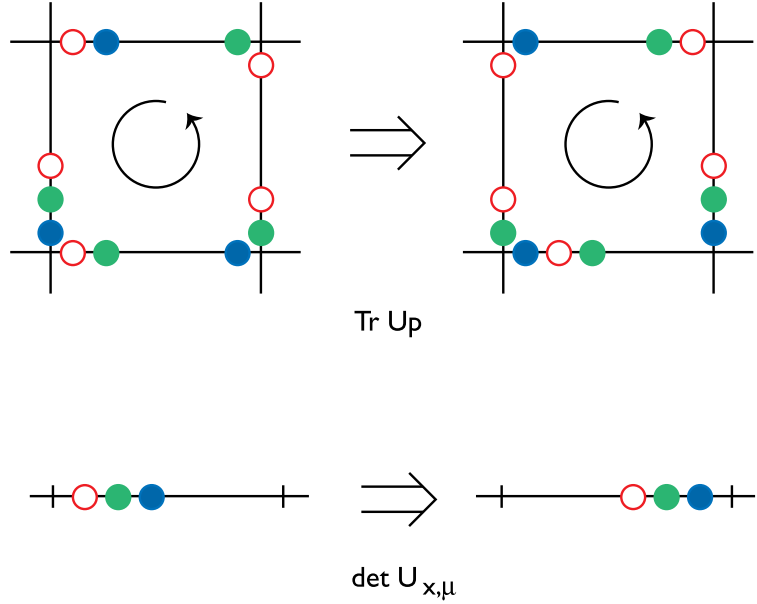,height=80.0mm}
\caption{Rishon dynamics: The trace part of the Hamiltonian induces a 
hopping of rishons of various colors around a plaquette. The
determinant part shifts a color-neutral rishon baryon from one end of
a link to the other.\label{fig:abacus}}
\end{figure}

The main goal of this talk is to display the rules for the quantum
link Hamiltonian in a form that allows one to appreciate the algebraic
structure in a vivid and intuitively appealing form. As illustrated in
Fig. \ref{fig:abacus}, the gauge sector of the quantum link QCD
Hamiltonian acts on a discrete Fermionic Hilbert space, represented by
colored dots for occupied Fermionic states associated with each
link. The link occupation level is set to half filling for the sum of
left triplet $(3)$ and right anti-triplet $(\bar3)$ link Fermions. The
plaquette term in the Hamilton operator slides the beads in the Abacus
along the link in the clockwise or anti-clockwise direction changing
the flux by one quantum. Gauge invariance demands that the same color
that leaves a site must move into that site from the adjacent link in
the plaquette, conserving rishon number for each color at the sites.
Another term is introduced into the Hamiltonian to remove the unwanted
U(1) gauge invariance which creates and destroys rishon ``baryons'' at
opposite ends of each link, reducing the gauge group from U(3) to
SU(3).  Once the origin of these rules is explained constructing other
quantum link Hamiltonians becomes a natural game of inventing
appropriate ``cellular automata'' rules on the Quantum Abacus.

At first it may seem hopeless or even paradoxical that one might
represent exactly a continuous gauge group (or even a global Lie
group) with a finite set of bits per lattice site. However, as
discussed in more detail in Wiese`s talk, a quantum spin is just such an
object.  For example the electron at rest has only 2 states {\bf
but} still respects exact rotational invariance via quantum
superposition.  Consequently the trick is to consider the appropriate
quantum operators in a discrete Hilbert space. Just as in all quantum
systems, the eigenvalues for sets of commuting operators enumerate the
states and the operators for Lie group symmetries ``rotate'' them
continuously into linear combinations of these eigenstates.

\subsection{U(1) Warm Up Exercise}

Before considering the full problem of non-Abelian gauge theory, 
it is instructive to begin with the simpler compact U(1)
Abelian gauge theory. With hindsight many of the basic concepts needed
for full QCD can be introduced.  Since we want to construct a 4-d gauge
Hamiltonian, it is useful (again with the benefit of hindsight) to
consider a Kogut-Suskind Hamiltonian~\footnote{If we had considered the
Kogut-Suskind Hamiltonian for a non-Abelian theory, we would have had
to distinguish between the electric fields that act as {\em right}
generators, $\hat E_R$, and those that act as {\em left} generators, $\hat E_L = \hat
U\da \hat E_R \hat U\da$.} in 4-d  as an example,
\be
H_{KS} = \frac{e^2}{2} \sum_{x,\mu} \hat{E}_{x,\mu} \hat{E}_{x,\mu} + 
\sum_{x,\mu \ne \nu} \frac{1}{2 e^2} \hat{U}_{x,\mu} \hat{U}_{x+\mu,\nu} 
\hat{U}\da_{x+\nu,\mu} \hat{U}\da_{x,\nu} \; .
\ee
The sums extend over lattice sites $x$ and directions $\mu,\nu = 1,2,3,4$.
Local gauge invariance {\bf on each link} $(x,x+\mu)$ requires,
\be
[\hat E,\hat U] = \hat U  \;\;\;\;  \;\;\;\; [\hat E,\hat U\da] 
= - \hat U\da  \;\;\;\;  \;\;\;\; [\hat U, \hat U\da] = 0 \; ,
\label{eq:algebra}
\ee
where to simplify the notation the $x,\mu$ subscripts have been
dropped \footnote{ Including the link subscripts would lead to equations such as
$[\hat E_{x,\mu},\hat U_{y,\nu}] = \hat U_{x,\mu} \delta_{x,y}
\delta_{\mu,\nu}$,
throughout. Since all commutators for operators on different links are zero dropping
these indices should not cause confusion.}.

It is easy to prove (just as with the canonical commutation relations
$[\hat q, \hat p] = i$) that the only solutions to this algebra
(\ref{eq:algebra}) are infinite dimensional ``matrices''.  For
example, in a basis in which all the $\hat U$'s are diagonal (the
``field basis''), $\hat E$ is given by differentiation,
\be 
\hat U \rightarrow \E{i  \theta} \;\;\;\; ,  \;\;\;\; 
   \hat E \rightarrow - i \frac{ d}{ d \theta} \; ,
\ee
where $\theta = e A$.  On the other hand, if we allow the commutator
$[\hat U, \hat U\da]$ to be non-zero, finite dimensional
representations can be found. This amounts to a new quantum
prescription with the group manifold represented as a non-commutative
geometry.  For example in the smallest representation, the rest of the
algebra is easily satisfied by 2 by 2 sigma matrices with
\be
\hat U \rightarrow \half (\sigma_1 + i \sigma_2) \equiv \sigma^+ \rightarrow \
\left |  \begin{array}{cc} 
0 & 1\\
0 & 0
\end{array} \right |  \; ,  
\label{eq:sigp}
\ee
\be
\hat U\da \rightarrow \half (\sigma_1 - i \sigma_2) \equiv \sigma^- \rightarrow \ 
\left |  \begin{array}{cc} 
0 & 0\\
1 & 0
\end{array} \right | \; ,
\label{eq:sigm}
\ee
and the gauge generator,
\be
\hat E \rightarrow \half \sigma_3 \rightarrow \half \ 
\left |  \begin{array}{cc} 
1 & 0\\
0 & -1
\end{array} \right | \; .
\label{eq:sig3}
\ee

In a basis where the E field is diagonal (the ``electric basis''), we
may view this 2 by 2 representation as simply a truncation of the
infinite dimensional representation to a single unit of flux. To show
this, let us introduce a set of Fermionic creation/destruction
operators,
$$\{c_l,c\da_{l'}\} = \delta_{l,l'} \; ,$$
one for each Fourier mode $l = \pm \half, \pm \smfrac{3}{2},\pm \smfrac{5}{2},
\cdots, \pm \infty$,
and constrain the Hilbert space to 
$${\cal N} = \sum_l \; c\da_l c_l = 1 \; .$$
Now the $U(1)$ commutation relations (\ref{eq:algebra}) are satisfied
by the operators, 
\bea
\hat U
&=& \int \frac{d\theta}{2\pi}  \phi(\theta)\E{i \theta} \phi\da(\theta)
= \sum_{l} c_l c\da_{l+1} \; ,\\
\hat E &=& \int \frac{d\theta}{2\pi}  \phi\da(\theta) \frac{d}{i d\theta} \phi(\theta) = \sum_l \; l \; c\da_l c_l \;,
\eea
where $\phi(\theta) = \sum_l a_l \E{i l \theta}$.  It is now straight
forward to demonstrate that the sole consequence for the algebra of
truncating the flux ( i.e.  restricting the sum in $\phi(\theta) =
\sum_l a_l \E{i l \theta}$ to $|l| \le L$ ) is to introduce the non-vanishing
commutation relation,
$$[\hat U, \hat U\da] = \sum^{L-1}_{l=-L} ( c\da_{l} c_{l} - 
c\da_{l+1} c_{l+1}) \ne 0 \;\; . $$
It is important to contrast this truncation procedure in the
``electric basis'' relative to the more conventional approach of
replacing the group manifold by a finite set of angles $\theta_n = n
\Delta \theta$. In the latter, the continuous symmetry group
(i.e. $U(1)$) is, at best, replaced by an approximating subgroup
(i.e. $Z_N$), whereas by truncating to low dimensional representations
in the Fourier space the full symmetry group is left intact!

By taking the lowest non-trivial flux on a link ($l = \pm \half$), we
obtain the smallest representation,
\be
\hat U = a b\da \;\;\;\; \hat U\da = - a\da b \;\;\;\;
\hat R = b\da b  \;\;\;\; \hat L = a\da a
\ee
where $a \equiv c_{-\frac{1}{2}}, b \equiv c_\frac{1}{2}$.
These satisfy the commutators
\be
[\hat R, \hat U] = \hat U  \;\;\;, \;\;\;
[L, \hat U] = - \hat U  \; \; ,
[ \hat U, \hat U\da] = \hat R - \hat L \; \; .
\ee
In the 2 dimensional subspace with $a\da a + b\da b = 1$ and $\hat  E \equiv
(\hat  R -\hat L)/2$, this reproduces {\bf exactly} our earlier sigma
matrix construction in Eqs.~(\ref{eq:sigp}-\ref{eq:sig3}).

Let us summarize the situation. We can construct a U(1) gauge
invariant Hamiltonian acting on a 4-d lattice simply by replacing the
link phase factors, $\exp[\pm i \theta]$, by sigma matrices,
$\sigma^{\pm}$. The electric field term, being the value of the
quadratic Casimir for the lowest dimensional representation, is just an additive
constant and can be dropped. The resulting quantum link Hamiltonian for $U(1)$
is
\be
\hat H =  \sum_{x,\mu \neq \nu} \sigma^+_{x,\mu} \sigma^+_{x+\mu,\nu} 
\sigma^-_{x+\nu,\mu} \sigma^-_{x,\nu} 
\ee
and its partition function is
\be
Z = \mbox{Tr} \exp(- \beta \hat H).
\ee
We note that in this Abelian example, we have made  use of two equivalent
representation: one in terms of explicit (sigma) matrices and another
in terms of link Fermions (or rishons). The strategy in the next
section is to explore  both of these approaches to generalize this
construction to SU(N) Yang-Mills theory.

\subsection{U(3) Quantum Link Theory}

Let us consider a  quantum link Hamiltonian,
\begin{equation}
\label{Hamdet}
\hat H = \sum_{x,\mu \neq \nu} \mbox{Tr} [\hat U_{x,\mu} \hat U_{x+\hat\mu,\nu} 
\hat U_{x+\hat\nu,\mu}\da \hat U_{x,\nu}\da] \; + \; \cdots
\end{equation}
for the  non-Abelian field operators $\hat U^{ij}_{x,\mu}$. In this
Hamiltonian each plaquette is the trace of the matrix product in color
indices $i, j = 1,\cdots, N$, and a direct product of q number
matrix elements. The dots in Eq.~\ref{Hamdet} represent other
operators that will be needed for full QCD, for improved actions,
etc.

For SU(3) the local gauge algebra is realized as an $SU_R(3)
\times SU_L(3)$ symmetry for each link operator $\hat U$,
\be
\label{eq:GaugeAlg}
[\hat R_\alpha, \hat U_{ij}] = \hat U_{ik} \lambda^\alpha_{kj}   \;\;\;, \;\;\;
[\hat L_\alpha, \hat U_{ij}] = - \lambda^\alpha_{ik}\hat  U_{kj}\; \; ,
\ee
where for notational simplicity we again suppress the link indices
$x,\mu$ in the above formula: $\hat  U \equiv \hat  U_{x,\mu}$, $\hat  R\equiv
\hat  R_{x,\mu}$ and $\hat  L \equiv \hat  L_{x,\mu}$.  Note that each link in fact has
9 operators $\hat U_{ij}$ for $i, j = 1,2,3$ transforming as a $(3,\bar 3)$
representation~\footnote{A better notation for these tensors would be
to use upper and lower indices $\hat  U_i^j$ but the standard practice in
lattice gauge theory ignores this nicety.}.  All commutators involving
different links are zero.  The SU(3) generators obey the Lie algebra,
\be
[\hat R_\alpha, \hat R_\beta] = 2 i f_{\alpha \beta \gamma} \hat R_\gamma \;\;\;, \;\;\; 
[\hat L_\alpha, \hat L_\beta] = 2 i f_{\alpha \beta \gamma}\hat  L_\gamma \;\;\;, \;\;\; 
[\hat R_\alpha, \hat L_\beta] = 0 \; ,
\ee
where the Gell-Mann   matrices~\footnote{To avoid annoying factors of
$\smfrac{3}{2}$, I have absorbed this factor into redefining the
$\lambda$~'s and the f and d structure constants. This corresponds
nicely with the traditional convention for SU(2): $\sigma_\alpha
\sigma_\beta = \delta_{\alpha \beta} + i
\epsilon_{\alpha \beta \gamma} \sigma_\gamma$. }  satisfy the algebra,
\be
\lambda_\alpha \lambda_\beta = 
\delta_{\alpha \beta} + i  f_{\alpha \beta  \gamma} \lambda_\gamma+ d_{\alpha \beta  \gamma} \lambda_\gamma \; .
\ee
If we introduce infinitesimal generators,
\begin{equation}
\hat G^\alpha_x = \sum_\mu (\hat R^\alpha_{x-\hat\mu,\mu} +\hat
 L^\alpha_{x,\mu}) \; ,
\end{equation}
for local $SU(3)$ gauge transformations, it is trivial to demonstrate
that as a consequence of the above Lie algebra the single plaquette
Hamiltonian (\ref {eq:GaugeAlg}) does obey local gauge invariance. Thus we have a general
algebraic framework for constructing quantum link Hamiltonians with
local gauge invariance. Surely this framework is not unique, but given
the reasonable assumption that all non-zero commutators are confined
to operators on a single link, it is probably the most natural.

\subsection{ Minimal Solution}

As we will explain in more detail in the next section, there are many
solutions to this algebraic problem. Here we attempt to construct the
lowest dimensional non-trivial solution, which for U(N) gauge theory
will turn out to be built on the smallest irreducible representation
of SU(2N). This representation has dimension 2N.

We introduce the ansatz,
\be
\label{eq:minansatz}
\hat U_{i j} = \delta_{i j} \hat U_0 + i \lambda^\alpha_{i j} \hat U_\alpha = 
\delta_{i j} 
\bmat  
0 & {\bf 1}\\	
0 & 0\\
\emat
+ i \lambda^\alpha_{i j}
\bmat  
0 & -i \hat \Lambda_\alpha \\	
0 & 0\\
\emat
\ee
and
\be
\hat R_\alpha = 
\bmat
\hat \Lambda_\alpha & 0\\
0 & 0
\emat    \;\;\;, \;\;\; 
\hat L_\alpha = 
\bmat
0 & 0 \\
0 & \hat \Lambda_\alpha\\ 
\emat
\; .
\ee
It turns out that the gauge algebra (\ref{eq:GaugeAlg}) is satisfied,
if the q-number link operators $\hat  \Lambda_\alpha$ obey exactly the same
Gell-Mann algebra as the c-number generators $\lambda_\alpha$,
\be
\hat  \Lambda_\alpha  \hat\Lambda_\beta = 
 \delta_{\alpha \beta} + i  f_{\alpha \beta  \gamma} \hat\Lambda_\gamma
+ d_{\alpha \beta \gamma} \hat\Lambda_\gamma \; .
\ee
Thus the $\Lambda$'s may also be represented as 3 by 3 matrices for each
link.  To prove this, substitute the ansatz into
Eqs.~(\ref{eq:GaugeAlg}) to get the two relations,
\bea
\label{eq:relR}
\hat R_\alpha ({\bf 1} + \lambda_\beta \hat\Lambda_\beta) &=&   
({\bf 1} + \lambda_\beta \hat\Lambda_\beta) \lambda_\alpha \\
\label{eq:relL}
({\bf 1} + \lambda_\beta \hat\Lambda_\beta) L_\alpha  &=&    
\lambda_\alpha ({\bf 1} + \lambda_\beta \hat\Lambda_\beta). 
\eea
These are satisfied identically. For example expanding the first
relations (\ref{eq:relR})  above yields
\be
\hat\Lambda_\alpha + \lambda_\alpha + i f_{\alpha \beta \gamma} 
\lambda_\beta \hat\Lambda_\gamma 
+ d_{\alpha \beta \gamma} \lambda_\beta \hat \Lambda_\gamma  =
\lambda_\alpha + \hat\Lambda_\alpha + i f_{\beta \alpha \gamma} \hat\Lambda_\beta \lambda_\gamma 
+ d_{\beta \alpha \gamma} \hat\Lambda_\beta \lambda_\gamma \;\;.
\ee
Using only the condition that $f_{\alpha \beta \gamma} =
Tr[[\lambda_\alpha,\lambda_\beta],\lambda_\gamma]/6i$ and $d_{\alpha
\beta \gamma} = Tr[\{\lambda_\alpha,\lambda_\beta\},\lambda_\gamma]/6$
are totally anti-symmetric and symmetric tensors respectively, the
identity holds.  The second relation (\ref{eq:relL}) is the conjugate
of the first.  (Clearly the same construction will work for any Lie
group.)

In addition to the SU(3) gauge transforms discussed above, there is a pair
of left and right U(1) gauge generators,
\be
\hat  R_0 = 
\bmat
{\bf 1} & 0\\
0 & 0
\emat    \;\;\;, \;\;\; 
\hat L_0 =
\bmat
0 & 0 \\
0& {\bf 1}\\ 
\emat
\; \; .
\ee
The sum, ${\cal N} = R_0 + L_0$, is a trivial symmetry since it
commutes with the link operators and the other generators, but the
difference, $T = R_0 - L_0$, leads to an additional (unwanted) gauge
symmetry,
\be
[\hat T, \hat U_{x,\mu}] = 2 \hat U_{x,\mu} \;\;.
\ee
The consequence of this extra $U(1)$ generator is that the standard
plaquette interaction in our Hamilton operator (\ref{Hamdet}) actually
represents a $U(3)$ gauge invariant Yang Mills theory.  Indeed because
of this unwanted symmetry, all earlier
attempts~\cite{Hor81,Orl90,Cha96} did not succeed in constructing a
Yang-Mills Hamiltonian suitable for QCD with only the SU(3) gauge
group.

One way to understand the problem is first to consider the standard
single plaquette ($Tr[ U_P]$) Wilson action and to note that this
expression is also invariant with respect to U(3) gauge
transformation.  Consequently, the restriction to SU(3) for the Wilson
theory is in fact introduced ``by hand'' by defining the path integral
with the Haar measure $dU$ restricted to the SU(3)
manifold. Therefore, in the quantum case the analogous procedure is to
insert a symmetry breaking operator into the trace similarly
restricting the phase space. A natural approach (which we will
ultimately implement) would be to add to the Hamilton operator a term,
$J' Det[\hat U_{x,\mu}]$, for each link. But in our present minimal
construction, the direct product of a link with itself vanishes (
$\hat U^{i j} \hat U^{k l} = 0$), so such a determinant term vanishes
identical as an operator! The problem can be traced to the fact that
the smallest representation just does not allow one to squeeze more
than one unit of flux onto a single link. One (rather inelegant ) way
out is to build the SU(3) determinant out of three different operators
spreading the flux over three different paths joining the point $x$ to
$x+\mu$. Fortunately, the rishon representation will afford a more elegant
and minimal solution to constructing the determinant locally on each
link.

\subsection{The Rishon Representation}

We now proceed to this  approach to constructing quantum link QCD
based on Fermionic triplets (or rishons), which naturally leads to higher
dimensional representations of SU(6). The basic algebraic structure of
these rishon models is  exactly the same as our minimal model given above. In
particular all the commutators (and therefore the group properties)
are unchanged.

In addition to the gauge commutators (\ref{eq:GaugeAlg}), let's
enumerate the other commutators.  Again the only non-zero commutators
for $\hat U_{x, \mu}$ and $\hat U^\dagger_{x,\mu}$ involve the same
link $(x,x+
\mu$) so we suppress the link index in discussing the algebra. With $\hat U
= \hat U_0 + i \lambda_\alpha \hat U_\alpha$ and $\hat U^\dagger = \hat U^\dagger_0 -i
\lambda_\alpha \hat U^\dagger_\alpha$, we have
\bea
[\hat  U_0, \hat U^\dagger_0 ] &=&    \hat T \; , \\ \relax
[ \hat U_\alpha, \hat U^\dagger_\beta ] &=& \delta_{\alpha \beta} \hat T
	+ i f_{\alpha \beta \gamma} (\hat R_\gamma + \hat L_\gamma)
         + d_{\alpha \beta \gamma} (\hat R_\gamma - \hat L_\gamma) \, \\ \relax
[ \hat U_0, \hat U^\dagger_\alpha ] &=& i (\hat R_\alpha - \hat L_\alpha).
\eea
Also the commutators,
\be
[\hat U_\alpha, \hat U_\beta] = 0\;\; , \;\; [\hat U\da_\alpha, \hat
U\da_\beta] = 0\; ,
\ee
are zero for all $\alpha,\beta = 0,\cdots N^2-1$. In addition in the minimal
construction above, we also have the nil potency property that $\hat U_{x, \mu}$
and $\hat U^\dagger_{x, \mu}$ just as in the U(1) gauge theory act
like a creation-destruction operators: $\hat U_{x, \mu}^2 = \hat
U^{\dagger 2}_{x, \mu} = 0$.   Not being part of the Lie algebra, we
will see that this special nil potency condition is a peculiarity of the lowest
dimensional representation.

The rishon construction begins with the  following ansatz.  Consider $2
N$ Fermionic operators $a_i$ and $b_i$ for $i = 1,\cdots ,N$, with
anti-commutators,
\be
\{a_i, a^\dagger_j \} = \delta_{i j}   \;\;\; , \;\;\;  \{b_i, b^\dagger_j \} = \delta_{i j} \; ,
\ee
on each link and {\bf all}  other  anti-commutators zero between different links and between a's
and b's on the same link.  The local gauge rotations are
\be
\hat L_\alpha = \sum_i a^\dagger_i \lambda^\alpha_{i j} a_j \;\;\; , \;\;\; \hat R_\alpha = \sum_i b^\dagger_i \lambda^\alpha_{i j} b_j \; ,
\ee
on each link with $\alpha = 0, 1 \cdots N^2 - 1$.  Just as in traditional
current algebras,  these bilinears obviously satisfies the correct  $SU_R(N)
\times SU_L(N)$ gauge
algebra.   The gauge fields are then  represented by
\be
\hat U_{i j} = \sum_{\alpha = 0}^{N^2 -1} \lambda^\alpha_{j i } \;  a_n\lambda^\alpha_{n m} b\da_m,
\ee
or 
\be
\hat U_{i j} =  a_i b\da_j \; ,
\ee
since the $N^2$ $\lambda$'s form a complete hermitian basis.
Again it is straight forward to see that the complete set of bilinears
made from $2 N$ Fermi fields must give the Lie Algebra
relations for $SU(2N)$. There is a very appealing interpretation. The
bosonic gauge field is a Fermion-anti-Fermion bound state. Like the tight binding
Hamiltonians of solid state physics, these ``rishons'' constituents never
move relative to each other, so there are no new internal degrees of freedom.


The rishon bilinears act on a $2^{2N}$ dimensional Fock space at each
link which can be enumerated by occupation numbers $0,1$ for each
Fermionic mode: $a\da_i,b\da_i$. Moreover the  entire algebra commutes with
the link number operator,
\be
{\cal N} = \hat R_0 + \hat  L_0 = \sum_i a^\dagger_i a_i + \sum_i b^\dagger_i b_i \; ,
\ee
so that the $2^{2N}$ dimensional matrices are reducible into a set of
$2 N +1$ irreducible representations.  The enumeration of the
representations  go as follows.  The link number $\hat R_0 + \hat L_0
= K$ extends over $K= 0, 1, \cdots 2 N$, but there is a  $Z_2$  particle-hole
symmetry that relates $K$ to $2N -K$ so really the models are
distinguished only for $K = 0,1,\cdots N$. The case K = 0 is trivial.
Consequently, there are only N distinct representations. For 
U(1) there is a unique model which is exactly the one discussed above
in Section 2.1.  For general N, the case K = 1 corresponds to a $2 N$
dimensional representation, since the Fock space is restricted to a
single Fermion with one of the $a_i$'s or one of the $b_i$'s occupied.
This is exactly the minimal U(N) construction presented above.  The
models with $K>1$ are new. For SU(2), K = 2 is a 6 dimensional
representation of SU(4) (anti-symmetric two index tensors). The square
of a link field now has one nonzero component given by
\be
det[\hat U_{x,\mu}] = \frac{1}{2!} \;\epsilon_{i j} \; \epsilon_{i' j'}\hat U^{i,i'}_{x,\mu} 
\hat U^{j,j'}_{x,\mu} = 2!\; a^{1 \dagger}_{x,\mu} a^{2 \dagger}_{x,\mu} b^1_{x,\mu} b^2_{x,\mu} .
\ee
Consequently a term which breaks U(2) down to SU(2) can be introduce
locally on a single link in the 6 dimensional representation.  For
SU(3) there are two new models --- for $K = 2$, a 15 dimensional
representations with $\hat U^2$ non-zero but $\hat U^3 = 0$ and for $K
= 2$, a 20 dimensional representations with $\hat U^3$ non-zero. Just
as in SU(2), the 20 dimensional representation allows a  U(1)
breaking term to be introduced on a single link,
\be
det[\hat U_{x,\mu}] = \frac{1}{3!} \; \epsilon_{i j k} \; \epsilon_{i' j' k'} \; \hat U^{i,i'}_{x,\mu}
\hat  U^{j,j'}_{x,\mu} \hat U^{k,k'}_{x,\mu} = 3! \; a^{ 1 \dagger}_{x,\mu} a^{2 \dagger}_{x,\mu} a^{ 3\dagger}_{x,\mu} b^1_{x,\mu} b^2_{x,\mu} b^3_{x,\mu} \; .
\ee
The intermediate case for SU(3) with K = 2 will allow U(1) breaking to
be introduced  on a single plaquette coupling a di-rishon with a
rishon. Clearly there is a tradeoff between the size of the
representation and the ability to represent operators locally on the
lattice.

In  summary, we  pick our candidate  Hamilton operator for SU(N) Yang-Mills theory to be 
\begin{eqnarray}
\hat H&=& J\sum_{x,\mu \neq \nu} \sum_{i,j,k,m} a^i_{x,\mu} b^{j \dagger}_{x,\mu} 
a^j_{x+\hat\mu,\nu} b^{k \dagger}_{x+\hat\mu,\nu} b^k_{x+\hat\nu,\mu} 
a^{m \dagger}_{x+\hat\nu,\mu} b^m_{x,\nu} a^{i \dagger}_{x,\nu} \\
&+& J' \sum_{x,\mu} \ N! \;  [a^1_{x,\mu} b^{1 \dagger}_{x,\mu} 
a^2_{x,\mu} b^{2 \dagger}_{x,\mu} ... a^N_{x,\mu} b^{N \dagger}_{x,\mu}
+ b^1_{x,\mu} a^{1 \dagger}_{x,\mu} 
b^2_{x,\mu} a^{2 \dagger}_{x,\mu} ... b^N_{x,\mu} a^{N \dagger}_{x,\mu}] \nonumber 
\end{eqnarray}
The plaquette part of the Hamilton operator shifts single rishons from
one end of a link to the other and simultaneously changes their
color. The determinant part, on the other hand, shifts an entire
``rishon-baryon'' --- a color neutral combination of $N$ rishons ---
along the link. Each link is restricted to a half filled state of N
rishons by fixing,
$$a^{ i \dagger}_{x,\mu} a^ i_{x,\mu} + b^{ i \dagger}_{x,\mu} b^i_{x,\mu} = N \; ,$$
so that the link algebra carries a $ 2 N !/(N! N!)$
dimensional irreducible representation. This rather simple rishon
dynamics may facilitate new analytic approaches to QCD, and may also
be useful in numerical simulations of quantum link models as discussed
in Section 3.

\subsection{Comments on SU(2) Gauge Actions}

SU(2) was the first example of SU(N) Yang Mills theory formulated as a
quantum link model~\cite{Hor81}.  For SU(2) there is a special
simplicity, which allowed the link operators $\hat U_\mu$ to be
represented by $\gamma_\mu$ matrices in an SO(5) algebra, but this
special feature tended to obscured further generalizations.  With our
general rishon method for SU(N), it is natural to ask if this
special SO(5) construction for the SU(2) model represents a distinct
alternative. In fact this is not the case. Following the rishon
construction, one can develop the SO(5) algebra for SU(2) as a
limiting case.

Returning to our minimal construction presented in Sec. 2.3, the
algebraic distinction between SU(2) and SU(N) is due to the d-symbol.
To remove the extra U(1) gauge invariance for SU(2), one makes use of
another form of the link operator, $\hat V_{ij} = (\sigma_2\hat
U^\dagger \sigma_2)_{ji}$ that also obeys the same gauge algebra for
SU(2), where $\hat U$ is given by the  expression in
Eq.~\ref{eq:minansatz}. This is possible for SU(2) because there is an
extra G parity symmetry: $\sigma_\alpha
\rightarrow \sigma_2 \sigma_\alpha \sigma_2 = - \sigma^*_\alpha =
-\sigma_\alpha^T$ that preserves the algebra. (Note invariance of
$\sigma_\alpha \sigma_\beta = \delta_{\alpha \beta} + i
\epsilon_{\alpha \beta \gamma} \sigma_\gamma$ under $\sigma_\alpha 
\rightarrow -(\sigma_\alpha)^T$.) In other words
for SU(2), the $(2)$ and $(\bar 2)$ representations, which are related
by the raising and lowering symbols ( $\epsilon_{ij} = \epsilon^{ij} =
i (\sigma_2)_{ij}$), are unitarily equivalent.

However if we attempt the same construction for SU(3) by introducing,
\be
\hat V = 
\bmat  
0 & 0\\	
{\bf 1} & 0\\
\emat
+ i \lambda_\alpha
\bmat  
0 &  0 \\
i \hat \Lambda_\alpha  & 0\\
\emat
\;\; ,
\ee
the new fields fail to obey the gauge algebra (\ref{eq:GaugeAlg}),  
\bea
- ({\bf 1} - \lambda_\beta \hat \Lambda_\beta) \hat R_\alpha & \ne&
({\bf 1} - \lambda_\beta \hat \Lambda_\beta) \lambda_\alpha  \;\; ,\\
\hat L_\alpha ({\bf 1} - \lambda_\beta \hat \Lambda_\beta)    & \ne&
-\lambda_\alpha ({\bf 1} - \lambda_\beta \hat \Lambda_\beta) \;\; ,
\eea
due to the opposite sign for the d-symbols on the right and left hand
sides. Consequently only for SU(2), do we have the special
construction,
\be
\label{eq:su2}
\hat H_{SU(2)} = \frac{J}{4} \sum_{x; \mu > \nu} \mbox{Tr} [{\cal U}_{x,\mu} 
{\cal U}_{x+\hat \mu,\nu} {\cal U}_{x+\hat\nu,\mu}\da {\cal U}_{x,\nu}\da] \; ,
\ee
where the new link operators are a sum of two terms, ${\cal U}_{x,\mu} =
\hat U_{x,\mu} + \hat V_{x,\mu}$. Note also that $U = {\cal U}
(1+T)/2$ and $V = {\cal U} (1-T)/2$. This Hamiltonian is ``by
accident'' both hermitian and gauge invariant under SU(2) but not
U(2).

However  another construction is also possible for SU(3), beginning
with the minimal 6 dimensional representation in
Eq.~\ref{eq:minansatz}. This is based on the observation that any link
$\hat U(x,y)$ connecting $x$ to $y$, transforms as a $(3,\bar 3)$
representation of $U_R(3) \times U_L(3)$. Thus not only can we form
U(3) gauge singlets by contractions with $\delta_{ij}$'s to make
``mesons'' at the sites but we can also form SU(3) singlets by
contractions with $\epsilon_{i j k}$'s to make ``baryons'' at the
sites.  For example we can add a term,
\be
\hat H_1 = J_1 \sum_{<x,y>} \epsilon_{i j k} \; \hat U^{(1)}_{i,i'}(x,y)
\hat U^{(2)}_{j,j'}(x,y) \hat U^{(3)}_{k,k'}(x,y) \; \epsilon_{i' j' k'}
\ee
where the $U's$ are 3 independent paths between x and y.

In the special case of U(2), one may use two paths connecting $x$ and
$y$ on a single plaquette. With appropriate weights and paths, this
procedure can lead you back to the earlier SO(5) suggestions for an SU(2)
Hamiltonians (\ref{eq:su2}). For example, take
\be
\hat H_1 = J_1 \sum_{<x,y>}\epsilon_{i j} \; \hat U^{(1)}_{i,i'}(x,y)  \hat U^{(2)}_{j,j'}(x,y) \;
\epsilon_{i' j'} \; .
\ee
Making use of the raising and lowering symbol $ \epsilon_{i j} = (i
\sigma_2)_{i j}$, this expression can be re-written as
\be
\hat H_1 = J_1 \sum_{<x,y>} \hat U^{(1)}_{i j}(x,y) \hat V^{(2)}_{j,i}(y,x)
\ee
Identifying $\hat U^{(1)}$ and $\hat V^{(2)}$ with products of links on the
edges of the plaquette, we get the individual terms in
Eq.~\ref{eq:su2} that break U(1) gauge symmetry. The
former choice (\ref{eq:su2}) is just one possible Hamiltonian, where
the epsilon symbol is inserted or not into all the corners of the
plaquette with equal weight.

\section{Quark Fields and Full QCD }

To represent full QCD, one must add quarks to the quantum link model.
This is more or less straightforward, although some subtleties arise
related to the dimensional reduction of Fermions~\cite{BroQCD}.

Again the guiding principle is to replace the classical action of the
standard formulation by a Hamilton operator which describes the
evolution of the system in a fifth Euclidean direction. For the quarks
this implies replacing the Grassmann variables  $\psi_x,\bar\psi_x$ by $\Psi_x,\Psi\da_x \gamma_5$,
where $\Psi\da_x$ and $\Psi_x$ are quark creation and annihilation operators
with canonical anti-commutation relations
\begin{equation}
\{\Psi^{i a \alpha}_x,\Psi^{j b \beta \dagger}_y\} = 
\delta_{xy} \delta_{ij} \delta_{ab} \delta_{\alpha \beta}, \
\{\Psi^{i a \alpha}_x,\Psi^{j b \beta}_y\} = 
\{\Psi^{i a \alpha \dagger}_x,\Psi^{j b \beta \dagger}_y\} = 0,
\end{equation}
where $(i,j)$, $(a,b)$ and $(\alpha,\beta)$ are color, flavor and Dirac
indices, respectively.

The  full QCD quantum link Hamiltonian is now given by
\begin{eqnarray}
\label{QCDaction}
\hat H&=&J \sum_{x,\mu \neq \nu} \mbox{Tr} [\hat U_{x,\mu} \hat U_{x+\hat\mu,\nu} 
\hat U\da_{x+\hat\mu,\nu}\hat  U\da_{x,\nu}]
+ J' \sum_{x,\mu} \ [\mbox{det} \hat U_{x,\mu} + \mbox{det}\hat  U\da_{x,\mu}] 
\nonumber \\
&+&\half\sum_{x,\mu} \ [\Psi\da_x \gamma_5 \gamma_\mu \hat U_{x,\mu} 
\Psi_{x+\hat\mu} - \Psi\da_{x+\hat\mu} \gamma_5 \gamma_\mu  \hat U\da_{x,\mu}
\Psi_x] + \sum_x \Psi\da_x \gamma_5 {\cal M}\Psi_x \nonumber \\
&+&\frac{r}{2} \sum_{x,\mu} \ [2 \Psi\da_x \gamma_5 \Psi_x - \Psi\da_x
\gamma_5 \hat U_{x,\mu} \Psi_{x+\hat\mu} - \Psi\da_{x+\hat\mu}
\gamma_5 \hat U\da_{x,\mu} \Psi_x].
\end{eqnarray}
The generators of $SU(N)$ gauge transformations must include quark  bilinears,
\begin{equation}
 \hat G^\alpha_x = \sum_\mu (\hat  R^\alpha_{x-\hat\mu,\mu} + \hat  L^\alpha_{x,\mu}) +
\Psi\da_x  \lambda^\alpha \Psi_x \; ,
\end{equation}
but it is still  trivial  to  show that $\hat H$ commutes with all $\hat G^\alpha_x$.

To ensure the proper dimensional reduction of the quarks, their
boundary conditions in the fifth direction must be chosen
appropriately. The standard antiperiodic boundary conditions, which
are dictated by thermodynamics in the Euclidean time direction, would
lead to Matsubara modes, $p_5 = 2 \pi (n_5 +
\frac{1}{2})/\beta$. However for our 5-th time, this would lead to a
mass of $O(\beta)$ for the dimensionally reduced quark.  On the other
hand, the confinement physics of the induced 4-d gluon theory takes
place at a correlation length which  grows exponentially with
$\beta$. Therefore, quarks with antiperiodic boundary conditions in the
fifth direction would remain at the cut-off freezing out the
quarks and returning the dimensionally reduced theory to a quarkless pure
Yang-Mills theory. One possible solution is
obvious. Simply choose periodic boundary conditions for the quarks in
the fifth direction. This gives rise to a Matsubara mode, $p_5 = 0$,
that survives dimensional reduction. Since the extent of the fifth
direction has nothing to do with the inverse temperature (which is the
extent of the Euclidean time direction), one could indeed choose the
boundary condition in this way.

There is another perhaps more attractive possibility. The above
scenario with periodic boundary conditions for the quarks still suffers
from the same fine tuning problem as the original Wilson Fermion
method. The bare mass matrix  $\cal M$ would have to be adjusted very
carefully in order to reach the chiral limit. In practice this is a
serious problem for numerical simulations. This problem has been
solved very elegantly in Shamir's variant ~\cite{Sha93} of Kaplan's
Fermion proposal ~\cite{Kap92}. Kaplan studied the physics of a 5-d
system of Fermions, which is always vector-like, coupled to a 4-d
domain wall that manifests itself as a topological defect. The key
observation is that under these conditions a zero mode of the 5-d
Dirac operator appears as a bound state localized on the domain
wall. From the point of view of the 4-d domain wall, the zero mode
represents a massless chiral Fermion. The original idea was to
construct lattice actions for {\bf chiral} gauge theories in this way.

Subsequently, Shamir pointed out that the same mechanism can also be used
to solve the lattice fine tuning problem of the bare Fermion mass in
vector-like theories like QCD. He also suggested a variant of Kaplan's
method that has several technical advantages, and that turns out to
fit very naturally with the construction of quantum link QCD. In
quantum link models, we already have a fifth direction for reasons
unrelated to the chiral symmetry of Fermions. We will now
follow Shamir's proposal, and use the fifth direction to solve the
fine tuning problem that we would have with periodic boundary
conditions for the quarks. The essential technical simplification
compared to Kaplan's original proposal is that one now works with a
5-d slab of finite size $\beta$ with open boundary conditions for the
Fermions at the two sides. This geometry limits one to vector-like
theories, because now there are two zero modes --- one at each
boundary --- which correspond to one left and one right-handed Fermion
in four dimensions. This approach  fits naturally with our construction
of quantum link QCD. In particular, the evolution of the system in the
fifth direction is still governed by the Hamilton operator of
Eq.~\ref{QCDaction}. The only  difference relative to Wilson's
Fermion method is that now $r < 0$.

The partition function of the theory with open boundary conditions for
the quarks and with periodic boundary conditions for the gluons is
given by
\begin{equation}
Z = \mbox{Tr} \langle 0|\exp(- \beta H)|0\rangle.
\end{equation}
Here the trace extends only over the gluonic Hilbert space of the
quantum link model, thus implementing periodic boundary conditions for
the gluons.  The boundary conditions for the Fermions are realized by
taking the expectation value of $\exp(- \beta H)$ in the Fock state
$|0\rangle$, which is annihilated by all right-handed $\Psi_{R x}$ and
by all left-handed $\Psi\da_{L x}$ ~\cite{Fur95}. As a result, there
are no left-handed quarks at the boundary at $x_5 = 0$, and there are
no right-handed quarks at the boundary at $x_5 = \beta$. Of course,
unlike periodic or antiperiodic boundary conditions, open boundary
conditions for the Fermions break translation invariance in the fifth
direction. Through the interaction between quarks and gluons, this
breaking also affects the gluonic sector. We don't expect this to be
problematic, because we are only interested in the 4-d physics after
dimensional reduction. The only crucial feature is that both quarks
and gluons have zero modes that survive dimensional reduction.

There is one technical detail. We have argued before that the gluonic
correlation length is not truly infinite as long as $\beta$ is finite,
but --- due to confinement --- it is exponentially large. The same is
true for the quarks, but for a different reason. In fact,
already free quarks pick up an exponentially small mass due to
tunneling between the two boundaries, which mixes left-handed and
right-handed states, and thus breaks chiral symmetry explicitly.

\begin{figure}
\psfig{figure=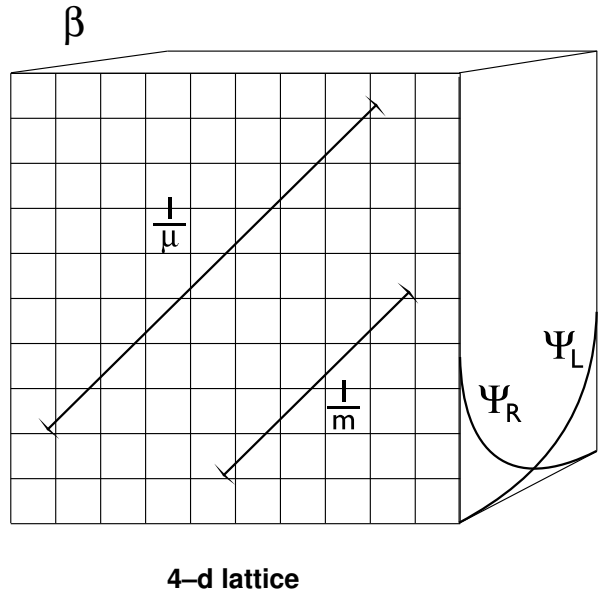,height=80.0mm}
\caption{Dimensional reduction in slab geometry: The gluonic correlation length
$1/m$ is exponential in $\beta$. Left and right-handed quarks live on the
two sides of the slab. The overlap of their wave functions $\Psi_R$ and
$\Psi_L$ induces a mass $\mu$. The corresponding correlation length $1/\mu$
is exponentially larger than $1/m$.\label{fig:slab}}
\end{figure}

The important observation is that the bare mass parameter $M$
(considering a single flavor for the moment) is not the physical
quark mass.  The physical mass of the quark goes exponentially fast
to zero in $\beta$ in the continuum limit without fine tuning.  At the
same time, the doubler modes remain at the cut-off.  Let us first
consider the doubler Fermions, which are characterized by $n =
1,2,3,4$ for the corners of the Brillouin zone.  It can be
shown~\cite{BroQCD} that the doubler  masses diverge, if we choose
\begin{equation}
M + 2 n r < 0.
\end{equation}
In fact this precise inequality applies for non-interacting quarks with
zero gauge fields. But close to the continuum limit, we expect nearly the
same inequalities.  Thus there is a  region with  $r < - M/2$, for a sufficiently
strong Wilson-term with an unconventional sign, where the doubler Fermions
are removed from the physical spectrum. On the other hand, the mass of
the physical Fermion (the $n = 0$ state) is
\begin{equation}
\mu = |E(\vec 0)| = 2 M \exp(- M \beta),
\end{equation}
which is exponentially small in $\beta$, again up to small corrections
when we turn on the gauge fields in weak coupling. These results
indicate how one may avoid the fine-tuning problem of the Fermion.
The confinement physics of quantum link QCD in the chiral limit takes
place at a length scale,
\begin{equation}
\xi \propto \exp(\frac{24 \pi^2 \beta}{(11 N - 2 N_f) N e^2}),
\end{equation} 
which is determined by the 1-loop coefficient of the $\beta$-function of QCD 
with $N_f$ massless quarks and by the 5-d gauge coupling $e$. As long as one
chooses
\begin{equation}
M > \frac{24\pi^2}{(11 N - 2 N_f) N e^2},
\end{equation}
the fictitious induced mass for the chiral quark is exponentially
smaller than the QCD scale $\Lambda_{QCD}$ and there is a window in
which the chiral limit is reached automatically as one approaches the
continuum limit taking $\beta$ large. For a given value of $r$ one is
limited by $M < - 2 r$ (note that $r < 0$). On the other hand, one can
always choose $J$ (and thus $e^2$) such that the above inequality is
satisfied.  The corresponding geometry is shown in
Fig. \ref{fig:slab}.

Of course, we also want to be able to work at non-zero quark masses. Following 
Shamir, we do this by modifying the boundary conditions for the quarks in the
fifth direction. Instead of using $\psi_L(\vec x,x_4,0) = 
\psi_R(\vec x,x_4,\beta) = 0$ we now demand
\begin{equation}
\label{boundary}
\psi_L(\vec x,x_4,0) = - \frac{m}{2 M} \psi_L(\vec x,x_4,\beta), \
\psi_R(\vec x,x_4,\beta) = - \frac{m}{2 M} \psi_R(\vec x,x_4,0),
\end{equation}
where $m$ is a mass parameter. This reduces to the previous boundary condition 
for $m = 0$, while it corresponds to antiperiodic boundary conditions for 
$m = 2 M$, and to periodic boundary conditions for $m = - 2 M$. Solving the 
above equations with the new boundary condition indeed yields physical quarks 
of mass $m$ in the continuum limit $\beta \rightarrow \infty$ as long as
$m \ll M$. It has been shown in Ref.~\ref{Fur95} that in the interacting theory
$m$ is only multiplicatively renormalized. On the level of the partition 
function, 
\begin{equation}
\label{partitionfunction}
Z = \mbox{Tr}[\exp(- \beta H) {\cal O}(m)] \; ,
\end{equation}
the new boundary condition manifests itself by a mass-dependent operator,
\begin{equation}
{\cal O}(m) = \prod_x 
(\Psi_{Rx} \Psi\da_{Rx} + \Psi\da_{Rx} \frac{m}{2 M} \Psi_{Rx})
(\Psi\da_{Lx} \Psi_{Lx} + \Psi_{Lx} \frac{m}{2 M} \Psi\da_{Lx}),
\end{equation}
which was constructed by Furman and Shamir~\cite{Fur95}. Note that in
Eq.~\ref{partitionfunction} the trace is both over the gluonic and
over the Fermionic Hilbert space. In the chiral limit $m = 0$, the
operator ${\cal O}(m)$ reduces to a projection operator $|0\rangle
\langle 0|$ on the Fock state introduced before. For $m = 2M$,
i.e. for antiperiodic boundary conditions, the operator ${\cal O}(m)$
becomes the unit operator, and the partition function reduces to the
 well-known expression from thermodynamics. For $m = -2M$, i.e. for
periodic boundary conditions, the operator ${\cal O}(m)$ becomes an
alternating sign $(-1)^F$ where $F$ is the total fermion number and
the trace becomes a super trace.

\section{Concluding Remarks}

In this brief talk, I have attempted to outline the  algebraic
steps required to  convert a standard lattice gauge theory into a
quantum link theory for the QCD Abacus. This has resulted in a
demonstration that we can in fact construct a 4-d quantum link
Hamiltonian for quarks and gluons which obey all the constraints of
local gauge invariance and discrete symmetries for QCD.  There has not
been time in this talk for many other important theoretical issues
such as why we believe this is a legitimate alternative formulation
of lattice QCD or time to report on the substantial progress being
made to convert our new scheme to a viable theoretical and
computational tool.  In a sense for quantum link QCD, we now have to
repeat the last 20 years of work that has been done in the Wilson
formulation.

Although it has not been emphasized, it should be plausible that our
general approach to converting a classical action into a corresponding
Hamiltonian operating on a discrete space in one extra dimension can
be applied to a wide range of quantum field theories.  In fact we have
coined the term ``D-theory'' for the general class of quantum spin and
quantum link theories that are obtained by such a discrete
``re-quantization'' of path integral expressions~\cite{Dtheory}.  In
some ways the procedure in D-theory is analogous to the construction
of M-theory whereby the string co-ordinates for the world sheet
embedded in 10 dimensions are replace by matrices living in 10+1
dimensions. When the extra dimension is compactified, the ratio
($e^2/\beta$) of the 11-d coupling ($e^2$) and the compact
circumference ($\beta$) becomes the coupling ($g^2$) in the 10-d
dimensionally reduce theory: $e^2/\beta \rightarrow g^2$. The general
procedure in D-theory is to study the symmetry properties of the
target field theory and to replace the co-ordinates (i.e. fields) with
a non-commuting algebra that expresses all the basic symmetries. Once
the representations for the fundamental quantum spins (matter fields) and
quantum links (gauge fields) have been chosen, a large variety of
Hamiltonians can be defined. If there is a zero gap in the d+1
dimensional D-theory before compactification of the extra dimension,
then we believe that dimensional reduction to and universality with
the target field theory is generally assured.

The use of anti-commuting fields to build finite dimensional
representations of symmetry groups (i.e. rishons in our terminology)
is of course a widely used device. Indeed the rishon formulation
resembles the Schwinger boson and constraint Fermion constructions of
quantum spin systems ~\cite{Aue94}. Just as it has for spins models
this trick may provide new theoretical insight leading perhaps to new
analytic approaches to QCD, for example, in the large $N$ limit. In
particular we note that the Hamilton operator~(\ref{QCDaction}) of
quantum link QCD with a $U(N)$ gauge group (i.e. dropping the
determinant term) can also be expressed in terms of color  singlet glueball, meson
and constituent quark operators.  These objects consist of two
rishons, two quarks and a quark-rishon pair, respectively. To see this,
just contract the color triplet indices in Eq.~\ref{QCDaction}, defining color singlet
bilinears for the new fields,
\be
\Phi_{x,+\mu,+\nu} =  c^{i \dagger}_{x,\mu} c^i_{x,\nu} \;\; , \;\;
M^{a \alpha b \beta}_x =  \Psi^{i a \alpha \dagger}_x \Psi^{i b \beta}_x  \;\; , \;\;
Q^{a \alpha}_{x,+\mu} =  c^{i \dagger}_{x,\mu} \Psi^{i a \alpha}_x  \; ,
\ee
representing glueball, meson and constituent quark mesons
respectively.  In this formula, we have defined $c^i_{x,\mu} =
a_{x,\mu}$ and $c^i_{x,-\mu} = b^i_{x,\mu}$ for $\mu = 1,2,3,4$. The
$8 + 4 N_f$ triplet fermion operators obviously combine as bilinears
to give the generators for an $SU(16 + 8 N_f)$ algebra.  In principle
this theory can be bosonized in terms of these degrees of freedom. If
large $N$ techniques can be applied successfully, major progress in
the non-perturbative solution of QCD and other interesting field
theories is to be expected. In any case, it is intriguing that $U(N)$
quantum link QCD has two representations --- one exclusively in terms
of colored quark and rishon Fermions, the other in terms of
color-singlet bosons. Returning to SU(N) only adds rishon baryons to
the set of color singlet fields.

Although it has not been emphasized in this talk, one can
argue~\cite{Cha96,BroQCD,WieseTalk} that there is a reasonable scenario
for this theory to undergo dimensional reduction so that the continuum
limit is reached for large values of the 5-th time extensions $\beta$ and
that the resulting theory is universally equivalent to Wilson's
formulation of QCD.  Let us remind ourselves of the analogous
situation for the conventional Wilson formulation.  Here the
``demonstration'' that lattice QCD gives the correct continuum theory
involved re-deriving renormalized perturbation theory from the lattice
and showing numerically that the confined phase extends to zero (bare)
coupling with asymptotically free scaling.  In spite of numerous
technical difficulties, there is a general consensus that these tests
have been met for the Wilson approach.  While we have not yet
accomplished a similar degree of certainty for the quantum link
formulation, we believe it can be accomplished. For Quantum link QCD,
perturbation theory is more difficult, but  we have begun to consider a
coherent state formalism which in principle will allow one to define
``spin waves'' (i.e. gluons) around a weak coupling vacuum. These spin
waves should interact in weak   coupling like ordinary QCD perturbation theory.

One approach is to begin with the equations of motion for the quantum link
Hamiltonian,
\bea
\frac{d}{dt} \hat U_{x, \mu} &=& i[\hat  H, \hat U_{x,\mu}] \;\; ,\;\;
\frac{d}{dt}\hat  T_{x,\mu} = i[\hat  H, \hat T_{x,\mu}] \; \nonumber\\
\frac{d}{dt}\hat  R^\alpha_{x,\mu} &=& i[\hat  H,\hat  R^\alpha_{x,\mu}]\;\; ,\;\;
\frac{d}{dt}\hat  L^\alpha_{x,\mu} = i[ \hat H,\hat  L^\alpha_{x,\mu}] \; .
\eea
and argue that in the appropriate classical limit, they lead as
expected to the 5-d Yang-Mills equation for the mean fluctuations ,
\be
A_\mu(x) = \frac{1}{2i}<\hat U(x+\mu,x) - \hat U^\dagger(x+\mu,x)> \;  ,
\ee
about the vacuum $<\hat U> = 1$ in the ``5-th time'' $A_5 = 0$ gauge.  In
addition, we are investigating the phase structure of 5-d QCD to
establish the existence of a gapless Coulomb
phase~\cite{Creutz,Chen}. The mechanism we postulate for dimensional
reduction of gauge theories is that compactifying the 5-th time leads
to a gap and confinement. As argued by Wiese and
Chandrasekharan~\cite{Cha96}, this follows closely an analogy with
dimensional reduction of the quantum spin Hamiltonians for 2-d
non-linear sigma models, where there is also a target asymptotically
free theory and the gap is guaranteed by the Mermin-Wagner-Coleman
theorem~\cite{MWC}.  Interestingly, in a non-Abelian gauge theory, the
analog of the Mermin-Wagner-Coleman theorem is confinement itself. The
direct demonstration of this mechanism for quantum link gauge theories
(since no one has succeeded to date in defining the confining vacuum)
will probably come from numerical simulations, where in the slab
geometry one identifies asymptotic freedom emerging as $\beta$ is
increased. This will require efficient Monte Carlo algorithms.

There is one last issue of tremendous promise and importance that must
be mentioned. In simulations, the potential advantage of the quantum
link approach is the application of cluster algorithms as typified by
the classic Swendsen-Wang algorithm~\cite{Swe87} that has
revolutionized simulations for the Ising model and other classical
spin systems~\cite{Wol89,Bro89}. In a cluster algorithm entire regions
of spatially correlated variables are ``flipped'' together. This can
have spectacular results near the continuum limit (or more generally a
second order phase transition) in reducing critical slowing down.  In
the standard Wilson formulation of gauge theories, no such cluster
algorithm has ever been found for models with a continuous gauge
group. For the U(1) quantum link model, we have succeed in formulating
and implementing the first such cluster algorithm.

\begin{figure}
\psfig{figure=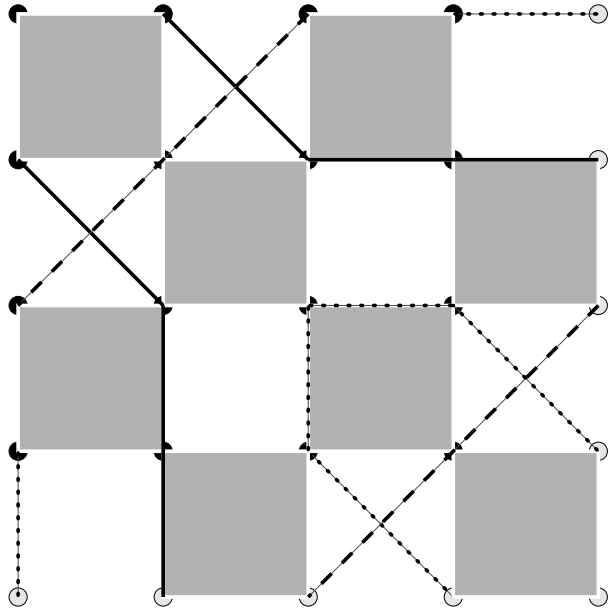,height=80.0mm}
\caption{ The figure shows a typical graph in 1d with $L=4$, and $N=2$. 
Every white square represents a two spin interaction. The graph shown
here splits the lattice into three clusters, represented by a solid
line, broken line and a dotted line. \label{fig.clust}}
\end{figure}

The cluster methods for the gauge model are being developed by a close
analogy with 2-d cluster algorithms for quantum spin systems.  For
quantum spin systems, successful cluster algorithms have already been
invented~\cite{Eve93,Wie94} and very impressive result obtain in
simulations~\cite{Bea96,Bea97a}.  For example efficient cluster algorithms
for a quantum XY models (analogous to a U(1) gauge theory) have been
studied and we know how to construct~\cite{Qclust} improved estimators
for quantum Green's functions such as $<\sigma^+_x \sigma^-_x>$.  By
use of an explicit basis for the transfer matrix and the Trotter
formula, one has a quantum generalization of the Fortuin-Kasteleyn map
onto a random cluster model.  The clusters for the quantum XY model
are single closed curves in the d+1 space (see Fig~\ref{fig.clust}).
We can prove detailed balance and basis independence for the quantum
matrix elements~\cite{Qclust}. Also one can formulate these algorithms
exactly on the continuous ``time'' interval $[0,\beta]$ making no
error due to the usual Trotter formula approximation.

Remarkably an analogous procedure has been carried out for the
U(1) gauge theory~\cite{Bas97}. The most interesting topological
difference with quantum spin system is that now the clusters are
closed surfaces (instead of closed lines) representing world sheets
for string states propagating in the extra time direction. As in all
efficient cluster algorithms with improved estimators, the structure
of the clusters should signify a new set of collective co-ordinates
which play an important role in the dynamics. For quantum link theory
these collective co-ordinates evidently are indicative of an underlying
stringy vacuum.

Finally we should remark that in the general formulation of a
particular D-theory, we expect that there are many alternative lattice
Hamiltonians that are universally equivalent after dimensional
reduction, but have different microscopic details. This flexibility at
the microscopic level is (ultimately) a strength. Just as solvable 2-d
spin systems, beginning with the classic work on the Ising model, have
led to new theoretical insight (i.e. critical scaling) and new
algorithms (i.e. percolation cluster methods), a particular
formulation of quantum link models may lead to more efficient
algorithms or more tractable analytic methods.  The implication is that
detailed choices among possible universally equivalent options should
go hand in hand with the development of analytical and numerical
methods.  There are many options that we are exploring that go beyond
the examples I have had time to present. Nonetheless we feel there is
a rather clear unified methodology in constructing a D-theory and
ample reason for optimism that they can contribute new insight into
the nature of asymptotically free field theories.

\section*{Acknowledgments}

I wish to gratefully acknowledge Shailesh Chandrasekharan and Uwe-Jens
Wiese for all aspects of this collaboration. They provide the ideal
circumstances where new ideas arise by a most enjoyable collective
effect. In addition there are many contributions, which are due to the
hard work of Maria Basler, Bernard Beard, Dong Chen, Kieran Holland
and Antonios Tsapalis.  I have also been stimulated in this research by
conversations with Samir Mathur on M-theory and Chung-I Tan on Matrix
Models. Finally I gratefully acknowledge the hospitality of the Center
for Theoretical Physics at Massachusetts Institute of Technology
during my frequent visits and to thank the organizers of AIJIC97 for
the opportunity to present this talk in an atmosphere most conducive
of a thoughtful exchange of ideas.

\eject


\vspace*{-9pt}
\section*{References}
\begin{thebibliography}{99}


\bibitem{Wil74}
K. Wilson, Phys. Rev. D10 (1974) 2445.

\bibitem{BroQCD} R. C. Brower, S. Chandrasekharan and U-J Wiese,
``QCD as a  Quantum Link Model'', hep-th/9704106.

\bibitem{WieseTalk} 
U-J Wiese {\em Quantum Spins and Quantum Links: From Antiferromagnets to QCD}
in this volume (1997).

\bibitem{Hor81}
D. Horn, Phys. Lett. 100B (1981) 149.

\bibitem{Orl90}
P. Orland and D. Rohrlich, Nucl. Phys. B338 (1990) 647.

\bibitem{Cha96}
S. Chandrasekharan and U.-J. Wiese, Nucl. Phys. B492 (1997) 455.

\bibitem{BroLinkFermion} R.~Brower, R.~Giles, and G.~Maturana.
``Link Fermions in Euclidean Lattice Gauge Theory'' {\em Phys. Rev.} (1984) D29:704.

\bibitem{Sha93}
Y. Shamir, Nucl. Phys. B406 (1993) 90.


\bibitem{Kap92}
D. B. Kaplan, Phys. Lett. B288 (1992) 342.


\bibitem{Fur95}
\label{Fur95} V. Furman and Y. Shamir, Nucl. Phys. 439 (1995) 54.

\ignore{ 
\bibitem{Blu96}
T. Blum and A. Soni, hep-lat/9611030. 

\bibitem{Har79}
H. Harari, Phys. Lett. 86B (1979) 83.
} 

\bibitem{Dtheory}
R. Brower, S. Chandrasekharan and U.-J. Wiese, in preparation.

\bibitem{Aue94}
A. Auerbach, ``Interacting Electrons and Quantum Magnetism'', Springer, 
New-York (1994).


\bibitem{Creutz} M. Creutz, Phys. Rev. Lett. 43  (1979) 553.

\bibitem{Chen} R. C. Brower, S. Chandrasekharan, D. Chen and U.-J Wiese, in preparation.


\bibitem{MWC} N. D. Mermin and H. Wagner, Phys. Rev. Lett. 17 (1966) 1133;
S. Coleman, Commun. Math. Phys. 331 (1973) 259.


\bibitem{Swe87}
R. Swendsen and S.-J. Wang, Phys. Rev. Lett. 58 (1987) 86.

\bibitem{Wol89}
U. Wolff, Phys. Rev. Lett. 62 (1989) 361; Nucl. Phys. B334 (1990) 581.

\bibitem{Bro89} R. C. Brower and P. Tamayo, Phys. Rev. Lett. 62 (1989) 1087


\bibitem{Eve93}
H. G. Evertz, G. Lana and M. Marcu, Phys. Rev. Lett. 70 (1993) 875.

\bibitem{Wie94}
U.-J. Wiese and H.-P. Ying, Z. Phys. B93 (1994) 147.


\bibitem{Bea96}
B. B. Beard and U.-J. Wiese, Phys. Rev. Lett. 77 (1996) 5130.


\bibitem{Bea97a}
B. B. Beard, A. Ferrando, M. Greven and U.-J. Wiese, in preparation.



\bibitem{Qclust}
R. Brower, S. Chandrasekharan and U.-J. Wiese, in preparation.

\bibitem{Bas97}
M. Basler, B. B. Beard, R. Brower, S. Chandrasekharan, A. Tsapalis and U.-J. Wiese, in 
preparation.

\ignore{ 

\bibitem{Has91}
P. Hasenfratz and F. Niedermayer, Phys. Lett. B268 (1991) 231.

\bibitem{Wie93}
U.-J. Wiese, Phys. Lett. B311 (1993) 235.

\bibitem{Gal96}
A. Galli, hep-lat/9605026.

\bibitem{Has93}
P. Hasenfratz and F. Niedermayer, Z. Phys. B92 (1993) 91.

\bibitem{Lue83}
M. L\"uscher, Commun. Math. Phys. 85 (1982) 39.

\bibitem{Nie96}
F. Niedermayer, Nucl. Phys. B (Proc. Suppl.) 53 (1997) 56.

} 

\end{thebibliography}
\end{document}